**Eliminating overgrowth effects in Poisson spatial process through the correlation among actual nuclei.**

M.Tomellini (*) M.Fanfoni (#)

(*) Dipartimento di Scienze e Tecnologie Chimiche Università di Tor Vergata and Istituto Nazionale per la Fisica della Materia, Via della Ricerca Scientifica 00133 Roma Italy

(#) Dipartimento di Fisica Università di Tor Vergata and Istituto Nazionale per la Fisica della Materia, Via della Ricerca Scientifica 00133 Roma Italy

**Abstract**

It has been shown that the KJMA (Kolmogorov-Johnson-Mehl-Avrami) solution of phase transition kinetics can be set as a problem of correlated nucleation [Phys.Rev.B**65**, 172301 (2002)]. In this paper the equivalence between the standard solution and the approach that makes use of the actual nucleation rate, i.e. that takes into account spatial correlation among nuclei and/or grains, is shown by a direct calculation in case of linear growth and constant nucleation rate. As a consequence, the intrinsic limit of KJMA theory due to the phenomenon of phantom overgrowth is, at last, overcome. This means that thanks to this new approach it is possible, for instance, to describe phase transition governed by diffusion.

In a recent paper we gave the exact solution of the problem of nucleation and growth in case of correlated nuclei [1,2], the correlation being due to the existence of an area around each island where nucleation is inhibited: it is customary to refer to such a zone as exclusion zone or capture zone. Previously we had discussed in a certain extend how the poissonian process, known as Kolmogorov-Johnson-Mehl-Avrami (KJMA) model [3,4,5], is a correlated problem itself when the actual nucleation is taken into account [6]. In fact an actual nucleus, by definition, cannot nucleate into a portion of the surface already transformed by previously formed nuclei. In this paper we go beyond mere, although reasonable, considerations, showing the equivalence between the two viewpoints by a direct calculation. Moreover, the kinetics, at odds with the KJMA theory, is computed on the ground of the actual (measurable) nucleation rate and this provides the solutions to the following long debated problems i) the applicability of KJMA to phase transitions governed by diffusion; ii) phase transformations kinetics where nucleation is prevented into a region of thickness around growing islands, as typical in second phase precipitation.

In particular, we shall take into account the case of constant nucleation rate throughout the whole space, $I$, that in the following will be referred to as virtual nucleation, to which corresponds the actual nucleation $I_a(t) = I[1 - S(t)]$. $S(t)$ is the kinetics of covered surface that, in the KJMA model can be expressed as

$$S(t) = 1 - e^{-Se_p(t)} \qquad (1)$$

and $Se_p(t)$ is the extended surface calculated on account of the virtual nucleation. The latter is computed considering square nuclei, because this choice allows to push the analytical calculation rather forward minimising, consequently, numerical computation. It is understood that all the squares have the same orientation so as to render inoperative the shielding effect [7,8]. Moreover, the growth law of the nucleus side has been taken linear, i.e. $\ell(t) = \upsilon(t - \tau)$, $t$ being the actual time, $\tau$ the time of its birth and $\upsilon$ is a constant. Eqn.1 becomes

$$S(t) = 1 - e^{-\int_0^t I\upsilon^2 (t-\tau)^2 d\tau} = 1 - e^{-kt^3/3} \qquad (2)$$

having defined $k = I\upsilon^2$.

The same result must be reached approaching the problem by using the actual nucleation function, that is introducing the spatial correlation among grains. As demonstrated in refs.1,2 the equation to be used is [1,2]

$$S(t) = 1 - e^{-\gamma(Se)Se} \qquad (3)$$

where $Se(t) = \int_0^t I[1 - S(\tau)]\upsilon^2 (t-\tau)^2 d\tau = k\int_0^t (t-\tau)^2 e^{-k\tau^3/3} d\tau$ is the extended surface of the actual nuclei, that is the total surfaces of actual nuclei regardless of impingement. The function $\gamma(Se)$ has been expressed by means of a series[1,2] of which we retained terms up to the second order. This implies the use of pair correlation functions for nuclei born at different times. To be specific, in the hard core model, the distances between the nuclei must satisfy the condition $d_{12} \geq \ell(\tau_2 - \tau_1)$, where $\tau_1 < \tau_2$ and $\ell(\tau_2 - \tau_1)$ is the size of the region around each nucleus where the nucleation is forbidden. In refs.1 and 2 has been shown that

$$\gamma = 1 + \frac{Se}{2} - \frac{1}{Se}\int_0^t I_a(\tau_2) d\tau_2 \int_0^{\tau_2} I_a(\tau_1)\Gamma[\ell(\tau_1,t),\ell(\tau_2,t)]d\tau_1 \qquad (4)$$

where

$$\Gamma[\ell_1,\ell_2] = \int_{-\ell_2/2}^{\ell_2/2} d\xi \int_{-(\ell_1/2+\xi)}^{\ell_1/2-\xi} dx \int_{-\ell_2/2}^{\ell_2/2} d\eta \int_{-(\ell_1/2+\eta)}^{\ell_1/2-\eta} dy\, W(x,y;\ell_{12}) \qquad (5)$$

In the latter equation $\ell_k = \ell(\tau_k,t)$, $\ell_{12} = \ell(\tau_2 - \tau_1)$ [see fig.1 for clarity] and $W(x,y;a)$ is the two dimensional square well of side $a$

$$W(x,y;a) =$$
$$= 1 - H\left(x+\frac{a}{2}\right)H\left(y+\frac{a}{2}\right) - H\left(x-\frac{a}{2}\right)H\left(y-\frac{a}{2}\right) + H\left(x-\frac{a}{2}\right)H\left(y+\frac{a}{2}\right) + H\left(x+\frac{a}{2}\right)H\left(y-\frac{a}{2}\right) \quad (6)$$

$H(x)$ is the Heaviside function and $W = 0$ for $-\frac{a}{2} < x < \frac{a}{2}$ and $-\frac{a}{2} < y < \frac{a}{2}$, i.e. no nucleation is allowed inside the square.

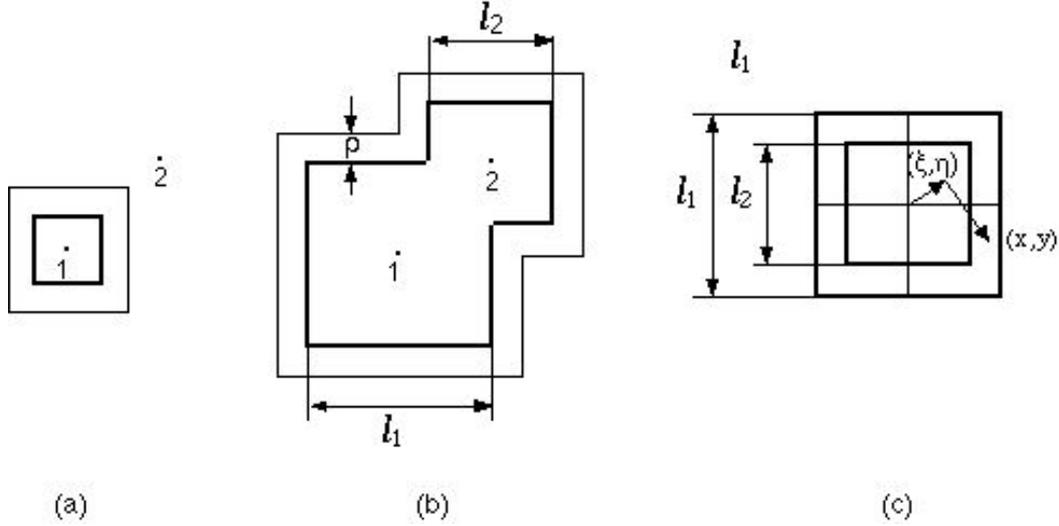

Fig.1 (a) Birth of nucleus 2 at time $t = \tau_2$. Its distance from nucleus 1, born at time $t = \tau_1 < \tau_2$, is larger than $\ell_{12} + 2\rho$. (b) The same nuclei as (a) at time $t > \tau_2$. (c) Integration domain for the 2D square well referred to (a) and (b). The integration domain of the variables $(\xi, \eta)$ is the square of side $l_2$, while the one of the variables $(x, y)$, whose are referred to the point $(\xi, \eta)$, is the square of side $l_1$.

In the following, for completeness, we shall treat the most general case, where we suppose the existence of the exclusion zone, which means that to the nucleus of side $\ell$ is associated an area precluted to nucleation equal to $(\ell + 2\rho)^2$ (fig.1) and then, from it, we shall compute the expression in keeping with the case under discussion $(\rho = 0)$.

Inserting eqn.6, with $a = \ell_{12} + 2\rho$, in eqn.5, an awkward but straightforward calculation leads to

$$\Gamma(\ell_1,\ell_2,\ell_{12};\rho)=4\left\{\frac{1}{2}\left(\omega+\frac{\ell_2}{2}\right)^2\left[\ell_1\ell_2-\frac{1}{2}\left(\omega+\frac{\ell_2}{2}\right)^2\right]H_1H_2+\omega\ell_2^2(\ell_1-\omega)H_3\right\} \qquad (7)$$

where $\omega=\left(\dfrac{\ell_1-\ell_{12}}{2}-\rho\right)$, $H_1=H\left(\omega+\dfrac{\ell_2}{2}\right)$ $H_2=H\left(-\omega+\dfrac{\ell_2}{2}\right)$ and $H_3=1-H_2$.

Restricting the calculation to the case of linear growth, $\ell_{12}=\ell_1-\ell_2$, and eqn.(7) becomes

$$\Gamma(\ell_1,\ell_2;\rho)=\left[2\ell_1\ell_2(\ell_2-\rho)^2-(\ell_2-\rho)^4\right]H(\ell_2-\rho). \qquad (8)$$

Moreover, since for the case under examination $\rho=0$, one ends up with

$$\Gamma(\ell_1,\ell_2)=\ell_2^3(2\ell_1-\ell_2). \qquad (9)$$

Eventually, thanks to eqns.(9) and (6), it is possible to numerically compute the kinetics eqn.(3).

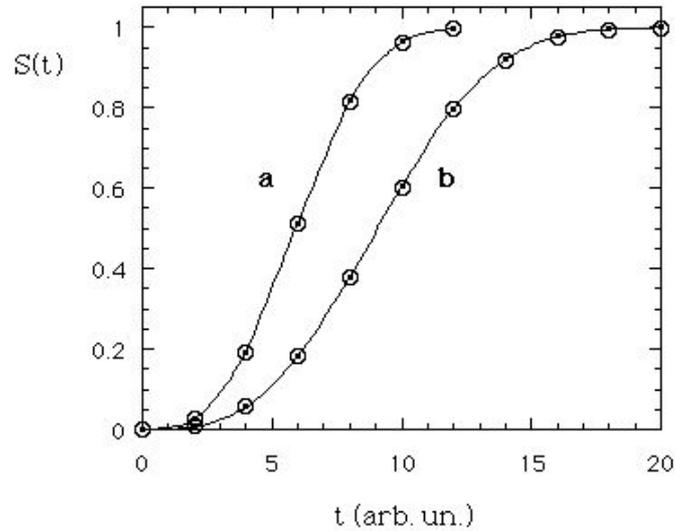

Fig.2 Nucleation and growth kinetics in case of random distribution of nuclei. Points: classical KJMA solution which implies the use of the virtual nucleation throughout the whole space in the computation of the extended volume (3D) or surface (2D). Open symbols: the kinetics are obtained by using the actual nucleation and, as a consequence, the use of the correlation functions. The parameters of the computations are: a) $k=10^{-2}$, b) $k=3.3\times10^{-3}$.

In fig.2 are displayed the kinetics obtained by using eqn.(1), which it is worth to remind is the exact solution, together with that obtained through eqn.(3). The latter is not exact just because the correlation is treated at the second order in the $g$-correlation functions [1,2], yet, as the comparison witnesses, it is doubtless more then good. The foregoing result is not a mere pedantic and in some way trivial remark -at least once determined how to manage the correlation- upon the KJMA model. On the contrary it tells us that the intrinsic limits of the KJMA formula can be overcome by the use of eqn.(3).

As a matter of fact, by employing the virtual nucleation can happen that some nuclei start growing in a fraction of space already covered by the new phase; these nuclei are known as phantoms [6,9]. They are not a problem whether the growth law of grains does not depend upon the size of the grains themselves in such a way that $\partial_t \ell(t-\tau_1) < \partial_t \ell(t-\tau_2)$, where $\tau_1 < \tau_2$. In this case, in fact, the covered younger grain will be able to overtake the older one giving rise to a non physical overgrowth event as shown in fig.3 [6,9,10]. In other words, there is a set of growth laws for which KJMA model is not applicable. This is not the case of eqn.(3) because it makes use of the actual nucleation, which, as an aside we note, is a measurable quantity.

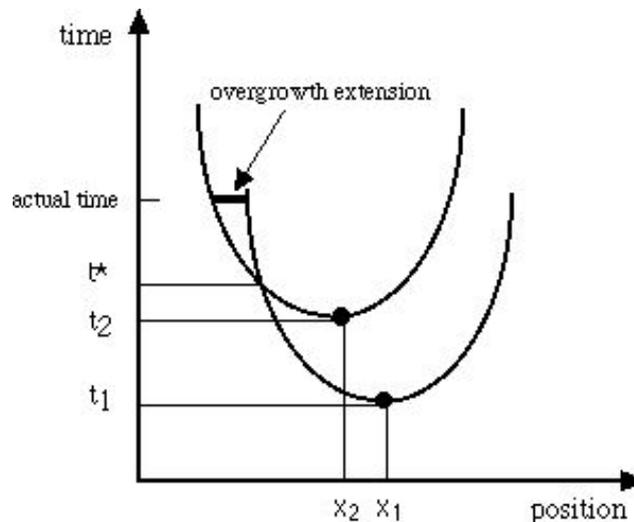

Fig.3 Overgrowth event in 1D of a phantom grain in case of parabolic growth law. The actual nucleus located at $x_1$ starts growing at time $t_1$ and the phantom nucleus located at $x_2$ starts growing at time $t_2$. The non physical overgrowth occurs at time $t^*$ at the intersection of the two parabolas. The segment that joins the two intersections between a parabola and a line parallel to the abscissa axis represents the grain size (1D in this case) at tha given time. The extention of the overgrowth is also highlighted at the actual time.

It is worth noting in passing that the formal solution of the surface (or volume) fraction kinetics, for any distribution of nuclei, was already proposed by Avrami in his first seminal paper [5] in terms of a series of the extended surface $Se_k$ as:

$$S(t) = \sum_k (-)^{k+1} Se_k. \tag{10}$$

The extended surface of order $k$ was expressed by Avrami as a multiple integral over the probability distribution function ($\tilde{f}$-functions), according to

$$Se_k = \int_0^t I_a(\tau_1) d\tau_1 \int_0^{\tau_1} I_a(\tau_2) d\tau_2 \ldots \int_0^{\tau_{k-1}} I_a(\tau_k) \overline{\upsilon}(\tau_1,\ldots,\tau_k) d\tau_k \tag{11}$$

where

$$\overline{\upsilon}(\tau_1,\ldots,\tau_k) = \int \upsilon(\tau_1,\ldots,\tau_k,\mathbf{x}_1,\ldots,\mathbf{x}_k) \tilde{f}_k d^k\mathbf{x} \tag{12}$$

and the quantity $\prod_i I_a(\tau_i) \tilde{f}(\mathbf{x}_1,\ldots,\mathbf{x}_k,\tau_1,\ldots,\tau_k) d\mathbf{x}_1\ldots d\mathbf{x}_k d\tau_1\ldots d\tau_k$ gives the probability of finding a nucleation center, which started growing between $\tau_1$ and $\tau_1 + d\tau_1$, in the element $d\mathbf{x}_1$ around $\mathbf{x}_1$, a nucleation center, which started growing between $\tau_2$ and $\tau_2 + d\tau_2$, in the element $d\mathbf{x}_2$ around $\mathbf{x}_2$ etc., irrespective of the location of the other nuclei. In eqn.(12) $\upsilon(\tau_1,\ldots,\tau_k,\mathbf{x}_1,\ldots,\mathbf{x}_k)$ is the overlap between $k$ nuclei (with birth time $\tau_1,\ldots,\tau_k$) located at $\mathbf{x}_1,\ldots,\mathbf{x}_k$. Avrami maintained that he would have approximated the kinetics by calculating the first two terms of the series (10), but never published the result, presumably because the approximation results rather poor. In this contest we stress that in eqn.(10) as many as infinite terms of the series contribute to a single term of the correlation function series in the exponent of eqn.(3). This is the reason why a second order truncation in the correlation function expansion leads to a very good agreement with the exact solution as displayed in fig.2. It is also interesting to note that once performed in eqn.(12) the cluster expansion of the

$\tilde{f}$-functions in terms of the $g$-correlation functions, the presence of the overlap volume does not permit to sum back the infinite $g$-series. As a matter of fact in refs.1,2 we showed that the surface fraction can be expressed as a different $f$-function series, with respect to eqn.(10), that, at variance with the latter, can be easily recast in terms of $g$-correlation functions.